# Effect of back-gate on contact resistance and on channel conductance in graphene-based field-effect transistors

A. Di Bartolomeo[1], S. Santandrea[1], F. Giubileo[2,*], F. Romeo[1,2], M. Petrosino[3], R. Citro[1,2], P. Barbara[4], G. Lupina[5], T. Schroeder[5], A. Rubino[3]

[1]Dipartimento di Fisica "E.R. Caianiello" and Centro Interdipartimentale NANO_MATES, Università di Salerno, via Ponte don Melillo, 84084, Fisciano (Sa), Italy

[2]CNR-SPIN Salerno, c/o Dipartimento di Fisica, Università di Salerno, via Ponte don Melillo, 84084, Fisciano (Sa), Italy

[3]Dipartimento di Ingegneria Elettronica ed Ingegneria Informatica, Università di Salerno, via Ponte don Melillo, 84084, Fisciano (Sa), Italy

[4]Physics Department, Georgetown University, Washington, DC 20057-1228, USA

[5]IHP, Im Technologiepark 25, 15236, Frankfurt (Oder), Germany

**Abstract**

We study the contact resistance and the transfer characteristics of back-gated field effect transistors of mono- and bi-layer graphene. We measure specific contact resistivity of ~7 k$\Omega \cdot \mu m^2$ and ~30k$\Omega \cdot \mu m^2$ for Ni and Ti, respectively. We show that the contact resistance is a significant contributor to the total source-to-drain resistance and it is modulated by the back-gate voltage. We measure transfer characteristics showing double dip feature that we explain as the effect of doping due to charge transfer from the contacts causing minimum density of states for graphene under the contacts and in the channel at different gate voltage.

[*]Corresponding author. Tel/Fax: +39 089 96-9329/5959. E-mail address:filippo.giubileo@spin.cnr.it (F. Giubileo)

# 1. Introduction

The need to ever improve the performances of field effect transistors (FETs) has made graphene an excellent candidate as channel material in nano-electronic applications. Better than carbon nanotube devices [1-2], graphene-based FETs (GFETs) [3] combine an ultra-thin body suitable for aggressive channel length scaling [4], with excellent intrinsic transport properties [5], such as the ability to tune the carrier concentration with electrical gates and a carrier mobility exceeding $10^4$ cm$^2$/Vs at room temperature [6]. GFETs with cut-off frequency of several hundred GHz have been demonstrated at room temperature [7]. For high performance, the choice of materials and fabrication techniques - both for the contacts [8] and the gate dielectrics [9-11] - play a very important role. In particular, the contacts between graphene and metal electrodes can significantly affect the electronic transport and limit or impede the full exploitation of the graphene intrinsic properties [12]. Although ohmic contacts are easily obtained on graphene due to the lack of a bandgap, the very small density-of-state of graphene around the Dirac point may suppress the current injection from the metal contacts, thus resulting in non-negligible contact resistance [12-13]. A high contact resistance $R_c$ limits the total on-state current, and has a detrimental impact on the peak transconductance and on the linearity of current versus gate-voltage characteristic [14], which are defining transistor parameters. It has also been shown that the current preferentially enters the graphene at the edge of the metal contacts, where current crowding can therefore take place [13]; accordingly an edge specific contact resistivity ($R_c W$ in Ω·µm, with $W$ width of the contact, see Fig. 1), is often preferred to a area specific contact resistivity $\rho_c$ ($\rho_c = R_c A$ in Ω·µm$^2$, with $A$ area of the contact) to characterize the contacts. Interface residuals, moistures, trapped charges, etc. may have high impact on the contact resistance and a careful control of the fabrication process is needed. Although in most experimental studies the effect of contact resistance on graphene devices can be suppressed by using a four-probe method, real applications require two-contact devices to achieve high integration. Therefore, understanding and controlling the contact resistance is important from a practical viewpoint other than from the basic physics related to the interface between a 3D metal and a 2D material. However, despite its importance, the contact resistance and its dependence on fabrication procedure and measurement conditions have not been yet exhaustively studied. Thus, it is still unclear why the reported values of $R_c$ span a very broad range. The lowest value of edge specific contact resistivity has been achieved for Pd (230±20 Ω·µm at room temperature [15]); for Ti leads, Ref.[16] quoted 0.8 kΩ·µm independent of the number of graphene layers but sensitive to the gate voltage and controllable by the pressure in the evaporation chamber (lower pressure giving lower resistance). Previous work on Ni leads [17] reported an area specific contact resistivity of few kΩ·µm$^2$ independent from the number of graphene layers. A smaller value, 0.5kΩ·µm$^2$, was reported in Ref. [13] for devices annealed in a $H_2$-Ar mixture. Robinson et al [8] propose a method based on $O_2$ plasma treatment of the graphene prior to metal evaporation to achieve area specific contact resistivity less than 10 Ω·µm$^2$ for any kind of metal, independently of their work function.

In this paper we study the contact resistance on mono- and bi-layer graphene sheets by fabricating structures suitable for transfer length method (TLM) measurements with Ni and Ti metals (materials of choice for micro/nanoelectronics industries) without any pre-metal treatment. Profiting of a relatively high specific contact resistance which results in the main contributor to the total source-drain resistance, we show that the contact resistance is modulated by the back-gate voltage ($V_g$). We further study source-drain conductance versus back-gate voltage, $G-V_g$ curves, and discuss the peculiar appearance of a double dip. We explain this feature by assuming that the graphene under the contacts is locally doped by the metal electrodes [18-19]. The charge neutrality condition for the graphene of the channel and that under the contacts is therefore achieved at a different values of $V_g$.

## 2. Sample preparation and measurement setup

Graphene flakes were obtained by exfoliation with adhesive tape from natural graphite (from NGS Naturgraphit GmbH) and transferred to Si/SiO$_2$ substrates. A short dip (~ 60s) in warm acetone was used to remove glue residuals. We selected several mono and bilayer graphene flakes to fabricated TLM structures, each one consisting of parallel leads (often of different length from 1 to 3 μm, see Fig. 1) at varying distances along the flake. The Si substrate is covered by a 300 nm thermal $SiO_2$ that allows easy optical identification of few-layer graphene. The number of layers was further confirmed by Raman spectroscopy and SEM imaging [20]. A grid of reference markers, with a pitch of 100 μm in x and y directions, was used to localize the graphene flakes. The markers were patterned by optical lithography and etched in the $SiO_2$ with a remaining oxide layer of about 100 nm.

The metal leads were patterned by electron beam lithography and standard lift-off in acetone at room temperature. Depending on the length of the flake, we deposited up to 6 contacts with inter-electrode distance ranging from 1 to 13 μm. The Ti and Ni metal leads were sputtered after keeping the sample in vacuum ($10^{-5}$ mbar) for about 24 hours to possibly remove adsorbates and mixtures from the graphene. We note that the work functions of Ti and Ni (4.33 eV for Ti and 5.01 eV for Ni) are respectively lower and higher than that of graphite (4.5 eV). The Ni and Ti leads were 70 nm thick and were coated with a 50-70 nm Au layer to prevent oxidation and to ease connection to the measurement probes.

SEM images of typical devices are shown in the inset of Figs. 1a) and 1b): the sample in Fig. 1a) is a monolayer contacted with Ni leads, while the sample of Fig. 1b) is a bilayer with Ti contacts (the bilayer is further exfoliated to become a monolayer under lead 4). Each couple of leads defines a FET with channel length (*L*) and width (*W*) and with back-gate formed by the highly p-doped silicon substrate. We chose to study devices with back-gate to avoid the risk that multiple-step processing – as necessary for a top gate – would impact the electrical properties of graphene.

Electrical measurements were performed at room temperature under ambient conditions with a Karl Suss probe station connected to a Keithley 4200 Semiconductor Parameter Analyzer. We performed 3-terminal measurements, with the Si substrate as the back-gate and all possible couples of metal electrodes as the source and drain. $V_g$ sweeps, in the interval (-80V, 80V), were performed at constant drain bias (typically 5 to 20 mV). Higher gate voltages were avoided to prevent oxide damage as stresses at $|V_g| > 80V$ systematically caused either gate leakage or complete oxide breakdown.

## 3. Results and discussion

*3.1 Contact resistance*

For an irregularly shaped sample (Fig. 1c), with contacts of different lengths, $d_1$ and $d_2$, and different widths, $W_1$ and $W_2$, and at a distance $L$, the two-probe resistance, $R = V/I$, is the sum of the channel resistance $R_{ch}$ and the contact resistances $R_{c1}$ and $R_{c2}$. Despite the current crowding effect at the contacts, it has been proven [13,21] that the contact resistance is characterized by the lead area (the area specific contact resistivity being $\rho_c = R_c W d$) rather than by the contact width ($W$) when the contact length ($d$) is shorter than the transfer length $d_T = \sqrt{\rho_c / R_{sh}}$, i.e. the effective length contributing to the current flow [13], which is typically longer than $1\mu m$ ($R_{sh}$ is the sheet resistance of the graphene channel in $\Omega/\square$). Since the typical length of most of our contacts is of the order of 1 $\mu m$, we decided to consider the area conduction rather than the edge conduction in our analysis. The total resistance of a trapezoidal graphene channel (see Fig. 1c) can be written as:

$$R = R_{c1} + R_{c2} + R_{ch} =$$
$$= \frac{\rho_{c1}}{W_1 d_1} + \frac{\rho_{c2}}{W_2 d_2} + R_{sh} \int_0^L \frac{dx}{W_1 + (W_2 - W_1)x/L} \quad (1)$$
$$= \frac{\rho_{c1}}{W_1 d_1} + \frac{\rho_{c2}}{W_2 d_2} + R_{sh} \frac{\ln(W_2/W_1)}{W_2 - W_1} L$$

If the contacts are made of the same metal, we can reasonably assume that $\rho_{c1} = \rho_{c2} = \rho_c$ and equation (1) yields

$$R_{eff} = \rho_c + \rho_{sh,eff} \frac{\ln(W_2/W_1)}{W_2 - W_1} L \quad (2)$$

where $R_{eff} = R \left( \frac{1}{W_1 d_1} + \frac{1}{W_2 d_2} \right)^{-1}$ and $\rho_{sh,eff} = \rho_{sh} \left( \frac{1}{W_1 d_1} + \frac{1}{W_2 d_2} \right)^{-1}$.

Equation (2) shows that the specific contact resistivity $\rho_c$ can be evaluated as the intercept of a plot of $R_{eff}$ versus $L$ for every couple of electrodes (TLM method). Examples are given in Figs. 1a) and 1b), where all the working combinations of electrodes were considered (some electrodes resulted in opens, some others were shorted by graphite flakes randomly present on the substrate). $R_{eff}$ was obtained from the I-V curves (output characteristics) measured at given back-gate voltages $V_g$ for all two-lead combinations on a flake. The output characteristics showed a linear behaviour both for Ni and Ti (an example is given in Fig 1d for Ni contacts), evidencing that enough good contact is established between graphene and metals.

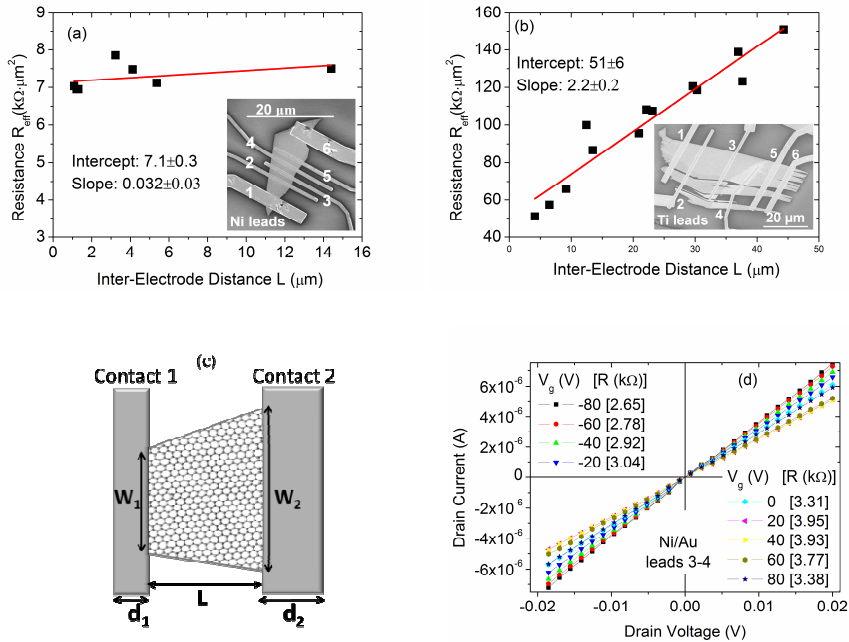

Figure 1. a) and b) TLM plot of $R_{eff}(L)$ at a given $V_g$ for the flakes reproduced in the insets; c) Schematic of a sample with varying channel width and with contacts of different lengths; d) Output characteristics of transistors of Fig. 1a) using Ni-leads 3 and 4 as source and drain.

Figures 2a) and 2b) show the dependence of $\rho_c$ on $V_g$ - both for Ni and Ti contacts - as extracted by the TLM method; $\rho_c$ is compared to the measured $R_{eff}(V_g)$ for the couple of leads which had the lowest resistance. These figures clearly demonstrate that the specific contact resistivity is modulated by the back-gate voltage. Furthermore, they show that the specific contact resistivity dependence on the back-gate voltage has the same qualitative behaviour as the total source-to-drain resistance (curve $R_{eff}(V_g)$), with a peak around the Dirac point ($V_{gD} \approx 20 \div 40V$) and a decrease when the channel is field-doped by the back gate. This result agrees with the recent study in Ref. [15] on Pd contacted graphene transistors and confirms the gate modulation of the Fermi level relative to the energy at the Dirac point for the graphene underneath the metal [22].

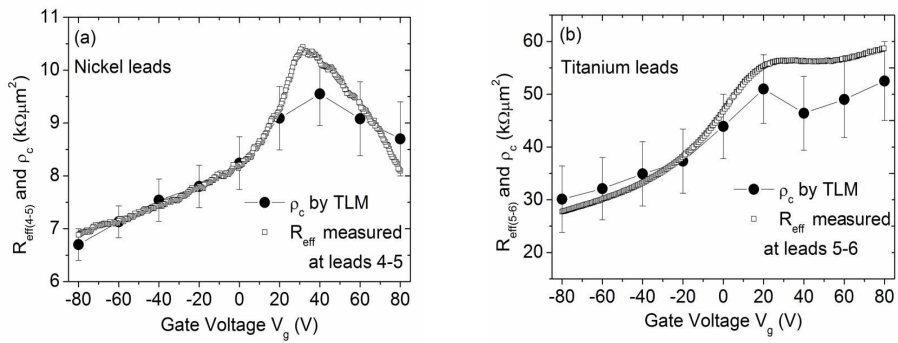

Figure 2: Contact resistivity $\rho_c$ and effective resistance $R_{eff}$ as function of the back gate potential $V_g$ for Ni and Ti contacts (devices of inset of Figs 1a and 1b, respectively).

Furthermore, the superposition of the two curves $\rho_c(V_g)$ and $R_{eff}(V_g)$ in the plots of Fig. 2 shows that the contact resistance is a major contributor to the resistance of the TLM transistors.

The minimum values of $\rho_c$ ~7 kΩ·µm$^2$ and ~30 kΩ·µm$^2$ we measured for Ni and Ti, respectively, are higher than those achieved by other groups; this difference can be ascribed to the fact that we did not make any pre-metal deposition treatment or any post-fabrication annealing of the device. Moreover, we deposited metal by sputtering and not by evaporation that usually results in lower contact resistance [8]. Additionally, we cannot exclude that some oxidation took place at contacts for oxygen diffusion and that we have somehow overestimated the effective contact area.

The flake of Fig 1b) allowed us to make a comparison between $R_{eff}$ when both leads are on bilayer graphene and in the case where one of the contacts is on a single layer (lead 4). No significant difference was observed (for example, $R_{eff} = 42\ k\Omega \cdot \mu m^2$ for leads 3-4 ad $R_{eff} = 44\ k\Omega \cdot \mu m^2$ for leads 3-5 at $V_g = -80V$), thus, confirming the independence of the contact resistance on the number of graphene layers reported by Venugopal et al [16]. Moreover, the conductance $G-V_g$ curves were similar.

*3.2 Conductance curves*

Figure 3 shows an anomaly in the conductance curves, appearing as a double-dip feature, that we often measured especially in long-channel GFET devices. We have already reported and extensively studied such feature for graphene transistors contacted with Cr/Au in a previous paper [18] where a wide hysteresis between the forward and reverse sweep and a higher inter-electrode distance created a clearer double-dip. The anomaly is observed both for Ni (Figs. 3a and 3b) and for Ti (Fig 3c), even though in the case of Ti, the limit on the back-gate sweeping range did not allow the full characterization of this phenomenon.

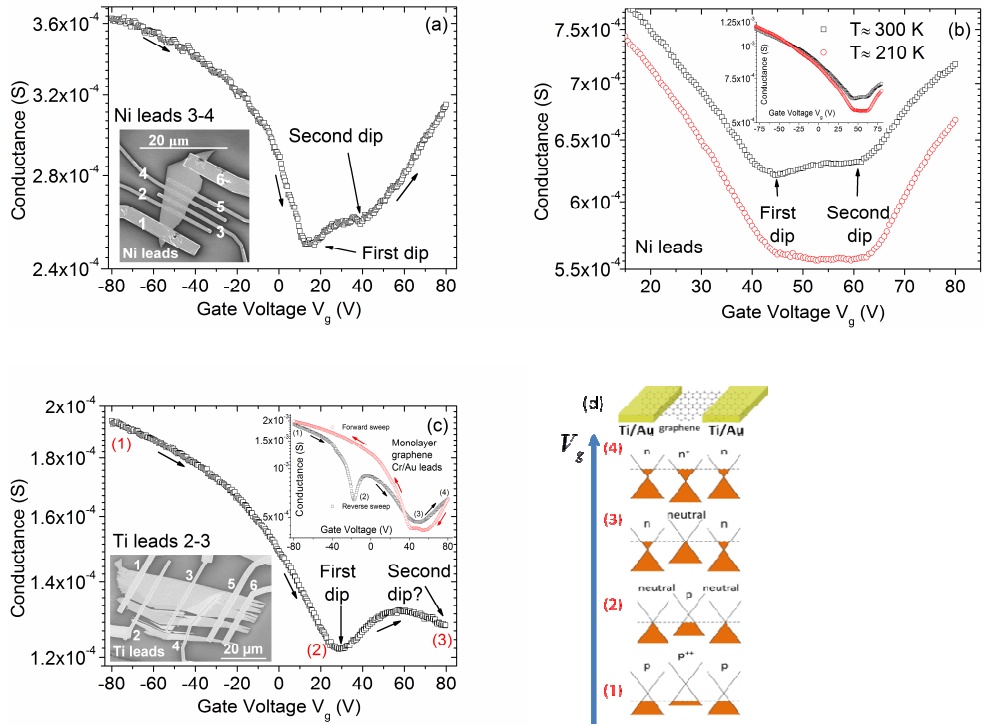

Figure 3: Appearance of a double dip. (a) and (b) conductance versus back-gate voltage curves for device with Ni contacts; (c) curves for device with Ti contacts. Inset of figure c) shows a $G-V_g$ loop for a monolayer graphene with Cr/Au leads discussed in Ref. [18]; (d) Model to explain the double dip feature.

The difference in work function between the metal and the graphene results in a charge transfer and doping of the graphene layer under the contacts [19]. The doping extends for 0.2–0.3 µm in the inner channel, making its effect barely detectable in shorter channel transistors. In longer devices metal doping yields two conductance minima at the energies of the Dirac points of graphene in the clamped and in the channel regions, which are observed as a double dip in the $G-V_g$ curve. In Ref. [18] we proved that metal doping and charge trapping at $SiO_2$/graphene interface, as well as partial pinning of the Fermi level at contacts, can fully explain the behavior of a whole $G-V_g$ loops like the one reproduced in the insert of Fig 3a and in particular, account for the double dip feature. Here, we qualitatively summarize the key points of our model with the help of Fig. 3d that shows a simplified band diagram from source to drain for a device with Ti leads, while forward sweeping the $V_g$ voltage. The electron transfer from Ti to graphene, due to the work function mismatch, makes graphene less p-doped underneath the contacts than in the channel. The application of the back-gate voltage moves the Fermi level in the graphene band diagrams, determining different conduction regions between source and drain. At $V_g = -80V$ (point 1 in figures 3c and 3d), p-type conduction takes place everywhere, thus giving a high conductance p/p$^+$/p structure (source/channel/drain graphene doping). While rising $V_g$ to positive values, electrons are attracted to the channel and a charge neutrality condition is reached at the contacts, where the graphene is less p-doped; a low conductance neutral/p/neutral structure is achieved, which corresponds to the first valley in the $G-V_g$ curve (point 2). A further increase in $V_g$ gradually reduces the p-doping in the channel and increases the n-doping at contacts creating an even less conductive n$^-$/p$^-$/n$^-$ structure (point 3). A further increase in $V_g$ leads to the formation of a low-resistance n-type (n/n$^+$/n) structure (point 4).

With Ni leads a similar behavior is obtained with the difference that graphene under the contacts is now more p-doped than that in the channel and a neutrality condition is first reached in the channel during the forward $V_g$, making the first minimum deeper than the second one.

Figure 3b shows that lowering the temperature leads to a slight reduction of the source-to-drain conductance that makes the two minima in the $G-V_g$ curve unresolved, the two dips being broadened at 210K compared to the expected V-shape. Two effects have been demonstrated to contribute to the broadening: the modification of the graphene density of states due to the coupling of graphene to metal *d*-states [23] and the formation of electron-hole puddles [24]. Further information can be extracted from Fig. 3b by observing that the minimum conductance and the carrier scattering life-time $\Gamma_T$ are related by the formula: $G^{\min}(T=210K)/G^{\min}(T=300K) = (210\Gamma_{300})/(300\Gamma_{210})$. Since the life-time broadening $\Gamma_T$ in graphene depends on the mean free path $\ell_{mfp}(T)$ through the phenomenological equation $\Gamma_T = (2\hbar v_F)/\ell_{mfp}(T)$, the experimental data imply $\ell_{mfp}(T=210) \approx 1.26\ell_{mfp}(T=300)$. Since the minimum of the conductance is proportional to the mean free path, $G^{\min}(T) \propto T \cdot \ell_{mfp}(T)$, its temperature scaling is consistent with what is expected for monolayer graphene [25].

However, despite the analysis of Figs 3a and 3b is fully consistent with the model given in Ref. [18], and although we have not observed a double deep in short channel devices (~1 μm), we cannot exclude that a double dip could arise in short-channel devices from a relevant spin-orbit effect. Indeed, as reported in Ref. [26], Ni-graphene interaction, especially if mediated by intercalation of Au atoms, produces an enhancement of the Rashba spin-orbit coupling which affects the conductance curves introducing cusp-like structures [27] similar to the one detected in our Fig 3a. The double dip mechanism described in Ref. [18] works for long-channel devices (channel length greater than a few μm), while in smaller GFETs double-dip-like structures in conductance curves could be produced by the strong proximity effects coming from the Ni-graphene interaction.

## 4. Conclusions

In conclusion, we have measured the area specific contact resistivity of Ti and Ni contacted graphene transistors. We have shown that, for untreated graphene, such resistivity is modulated by the back gate, with the same dependence as the channel resistance, i.e. a peak at the Dirac point and an asymmetric decrease for increasing $\left|V_g - V_{gD}\right|$. We have further shown and discussed a peculiarity, namely a double dip, of the $G - V_g$ curve that appears especially in longer transistors; we have explained such feature as corresponding to the two minima of the density of states of graphene respectively in the channel and under the metal, which are different because of the metal doping of graphene.